# Comparing Speech and Keyboard Text Entry for Short Messages in Two Languages on Touchscreen Phones

SHERRY RUAN, Stanford University
JACOB O. WOBBROCK, University of Washington
KENNY LIOU, Symantec Corp.[1]
ANDREW NG, Stanford University[1]
JAMES A. LANDAY, Stanford University

With the ubiquity of mobile touchscreen devices like smartphones, two widely used text entry methods have emerged: small touch-based keyboards and speech recognition. Although speech recognition has been available on desktop computers for years, it has continued to improve at a rapid pace, and it is currently unknown how today's modern speech recognizers compare to state-of-the-art mobile touch keyboards, which also have improved considerably since their inception. To discover both methods' "upper-bound performance," we evaluated them in English and Mandarin Chinese on an Apple iPhone 6 Plus in a laboratory setting. Our experiment was carried out using Baidu's Deep Speech 2, a deep learning-based speech recognition system, and the built-in QWERTY (English) or Pinyin (Mandarin) Apple iOS keyboards. We found that with speech recognition, the English input rate was 2.93 times faster (153 vs. 52 WPM), and the Mandarin Chinese input rate was 2.87 times faster (123 vs. 43 WPM) than the keyboard for short message transcription under laboratory conditions for both methods. Furthermore, although speech made fewer errors *during* entry (5.30% vs. 11.22% corrected error rate), it left slightly more errors in the *final* transcribed text (1.30% vs. 0.79% uncorrected error rate). Our results show that comparatively, under ideal conditions for both methods, upper-bound speech recognition performance has greatly improved compared to prior systems, and might see greater uptake in the future, although further study is required to quantify performance in non-laboratory settings for both methods.

## 1 INTRODUCTION

People today spend immense amounts of time texting using smartphones [40]. But a smartphone's small touch-based keyboard can be slow and frustrating to use. Given the amount of time people are spending on phones and other mobile devices, it remains important to design effective off-desktop text entry methods that can greatly reduce users' frustrations and improve efficiency [58]. Different text entry methods have been designed and implemented in recent years and extensive research has been conducted to evaluate their effectiveness in different settings [8,38,44,53]. Most methods are focused on virtual or physical keyboards or keypads, often with alternative key arrangements or letter disambiguation algorithms [26,56]. Speech has attracted some interest [34,44] and there are now several popular speech-based assistants, such as Apple Siri, Microsoft Cortana, Google Now, Amazon Echo Alexa, and Baidu Duer. But little is known about how these modern speech recognizers perform, even under ideal conditions, and how this performance compares to mobile touchscreen keyboards in similarly ideal settings. A study comparing the two would give us an "upper-bound" on performance from which to venture into further studies in less-than-ideal conditions more specific to the strengths and weaknesses of these and other methods.

---

[1] Authors K. Liou and A. Ng worked for Baidu USA when the study described in this article was planned, executed, and documented.

Despite decades of speech recognition research, speech recognition accuracy has not been sufficiently high for speech systems to enjoy widespread use. Indeed, back in 1999, Karat et al. [18] concluded that the accuracy of speech input was far inferior to desktop keyboard input. Technical constraints included ambient noise and the lack of support for out-of-vocabulary words [59]. Shneiderman [39] has argued that the same cognitive and memory resources that are used in speech production are also used in problem solving, hindering human performance in text composition, which might limit the acceptance of speech as a viable text input method.

However, in the last several years, there have been great advances in speech recognition due to the advent of deep learning models and advances in computational performance [15,35]. Indeed, speech recognition recently surpassed the threshold of having superior accuracy to human recognition, albeit in very limited contexts [1]. In light of these advances, it is now pertinent to re-explore the potential of speech-based text entry, specifically for input into today's smartphones and other mobile devices. We hypothesize that while speech recognition systems in the past might not have competed favorably with keyboard-based typing methods, today's state-of-the-art speech recognition systems might outperform smartphone keyboards. To test our hypothesis, we designed a touchscreen phone interface integrating state-of-the-art speech and keyboard input methods. We conducted a controlled experiment comparing the speech recognition system and the touch-based smartphone keyboard to evaluate the performance of the two under ideal laboratory conditions—a quiet indoor setting with participants seated at a table without distraction. By using the same ideal conditions and a text transcription task, rather than a text composition task, we isolated the performance of both state-of-the-art input methods to compare them with their "best foot forward." Future studies, then, can see how much non-ideal conditions degrade performance, or whether today's "upper-bound performance" can ever be achieved "in the wild." Given the popularity of touchscreen phones worldwide, we also compared text entry methods in both English and Mandarin Chinese, one of the only studies to do so.

In our study, we found that text entry speeds in words per minute (WPM) using speech were 2.9 times faster than the keyboard for English (153 vs. 52 WPM), and also about 2.9 times faster than the keyboard for Mandarin Chinese (123 vs. 43 WPM). Corrected error rates (i.e., errors made and fixed *during* entry) were also favorable to speech, with speech error rates being 16.7% lower than keyboard error rates in English (3.93% vs. 4.72%), and 62.4% lower in Mandarin (6.67% vs. 17.73%). However, speech input left slightly more errors than keyboard input *after* entry was completed in English (0.55% vs. 0.35%) and in Mandarin (2.06% vs. 1.22%). Thus, speech was demonstrably faster and more accurate than the keyboard *during* entry, but slightly more prone to leave errors *after* entry.

The chief contribution of this work is empirical: To the best of our knowledge, ours is the first rigorous evaluation of a state-of-the-art deep learning-based speech recognition system and a state-of-the-art touch-based keyboard for mobile text entry. Moreover, this contribution is made for two languages, English and Mandarin Chinese: the former is the most "influential" language worldwide, and the latter is the most widely spoken language worldwide [62]. In addition, we offer a design contribution: a new method of error correction that can utilize speech or the mobile keyboard. We also offer a methodological contribution: We report novel speech-specific measures that can be reused in subsequent evaluations of speech-based text entry. Finally, we offer insights for how to improve interaction designs for speech-based text entry.

For smartphone users wishing to have a more efficient text input mechanism, this research suggests that modern deep learning-based speech recognition systems might be an effective mechanism, although further research is warranted to test both speech and manual text input in less-than-ideal conditions (e.g., while walking, with ambient noise, with distraction, etc.). The "upper-bound performance" we establish here motivates future investigations and provides a point of comparison.

## 2 RELATED WORK

In this section, we mainly discuss studies applicable to speech-based text entry. The reader is directed elsewhere for a thorough review of text entry methods more generally [26,58,63].

Research has shown that humans' speaking rate can be as fast as 200 WPM for English [33] and 250 characters per minute (CPM) for Mandarin Chinese [54]. However, no study to date has claimed to achieve such text entry rates on mobile devices. In a user study of Parakeet [44], a continuous speech recognition system for mobile phones, participants entered text at an average rate of 18 WPM when seated and 13 WPM when walking. Another study [18] showed that users reached only 7.8 WPM for text composition and 13.6 WPM for text transcription using speech, compared to 32.5 WPM for a keyboard-mouse method. In contrast, users were able to achieve a higher entry rate with elaborately designed keyboards and some practice. A longitudinal study of a mini-QWERTY keyboard showed that participants reached an average of 60 WPM after 20 twenty-minute typing sessions [7]. There are scant rigorous research results on the performance of Mandarin Chinese speech or typing-based input methods.

Past research also reveals several limitations of speech recognition accuracy. In a study of data entry on the move, Price et al. [32] observed a recognition error rate of about 33-44%, and they concluded that this may be partly due to background noise, a common and persistent problem for deployed speech-based systems. Furthermore, Bradford [3] claimed that recognizing user actions with speech recognition was inherently error prone and no reliable solution to this problem existed. However, part of his reasoning was built upon a research result from 1988, when speech recognition systems were mostly based on signal processing and pattern matching, not deep learning [21].

The low accuracy of previous speech input methods might also be ascribed to the use of speech for error correction. In fact, correcting speech recognition errors *with speech commands* has been shown to be susceptible to cascading failures, in which correction commands are misinterpreted by the speech recognition system and themselves have to be corrected [18]. As with existing phone-based systems, we incorporate a touchscreen keyboard in our speech input method and provide the user with the flexibility to correct errors using either speech *or the keyboard* [43]. Our results reveal that most users prefer the keyboard to speech for correcting errors, and that this significantly improves performance.

Speech input methods have also been shown to be disliked by users. A longitudinal study showed that seven out of eight new users abandoned their speech recognition systems after six months, mainly due to their unsatisfying user experience with speech recognition [19]. Another review expressed users' concern for speech input because of its lack of privacy, security, and confidentiality in social settings [37].

Admittedly, results from previous speech input studies were not competitive compared to those of mobile keyboard input methods. However, speech recognition technology has made significant strides in recent years and, although the larger contextual and social factors surrounding speech recognition are not ameliorated by technical improvements, speech recognition accuracy has certainly improved. Recent advances arise in part because of the availability of large amounts of data, computation, and sophisticated deep learning models [1]. We expect that today's speech recognition systems have the potential to be suitable for general-purpose text entry. A first step is to quantify how modern speech systems perform compared to modern touch-based keyboards, both under ideal conditions, to establish their "upper-bound performance." To this end, we conducted the following experiment.

## 3 EXPERIMENT

To evaluate the "upper-bound performance" of two state-of-the-art mobile text input methods, speech recognition and typing on a touch-based keyboard, in two languages, English and Mandarin Chinese, we ran a controlled laboratory experiment. Our goal was not only to capture high-level measures such as text entry speed and accuracy, but also to reveal how a speech interface might be improved based on the low-level measurements we obtained.

### 3.1 Participants

A total of 48 people participated in this study. Twenty-four were native speakers of American English and 24 were native speakers of Mandarin Chinese. All participants were university students majoring in various fields including computer science, materials science, economics, chemistry, and business. Every participant was familiar with either an English QWERTY keyboard or a Mandarin Pinyin QWERTY keyboard on an Apple iPhone. Participants ranged in age from 19 to 32 years old ($M$=23.5, $SD$=4.1). Participants used both text input methods, keyboard and speech, only in their native language,

English or Mandarin Chinese. Twelve of the 24 English language participants were females and 12 were males, and the same ratio held for Mandarin participants. The experiment was conducted under the direct supervision of the first author. The study for each participant took about 30 minutes and participants received a small cash payment for their time. We performed the study in a quiet meeting room. Participants were seated at a table, not walking, and outside distractions were eliminated.[2]

## 3.2 Apparatus

We conducted our experiment on an Apple iPhone 6 Plus. We developed a custom experiment test-bed app using Swift 2 and Xcode 7, and connected this app to a state-of-the-art speech recognition system, Baidu Deep Speech 2 [1]. The speech recognition system ran entirely on a server off-site at Baidu. As we were connected to our university's high-speed network, there was no noticeable latency between the client iPhone and the speech server. Our test-bed app also utilized Apple's state-of-the-art built-in QWERTY keyboard for English and Pinyin QWERTY keyboard for Mandarin Chinese. Thus, two state-of-the-art commercial text entry methods were compared.

Our test-bed app presented phrases for transcription using two text input user interfaces: keyboard and speech. (Section 3.4 discusses the text entry phrase set and rationale for text transcription, instead of text composition.) Fig. 1(a) and (b) show the keyboard input interfaces with the English QWERTY and Pinyin QWERTY keyboards.

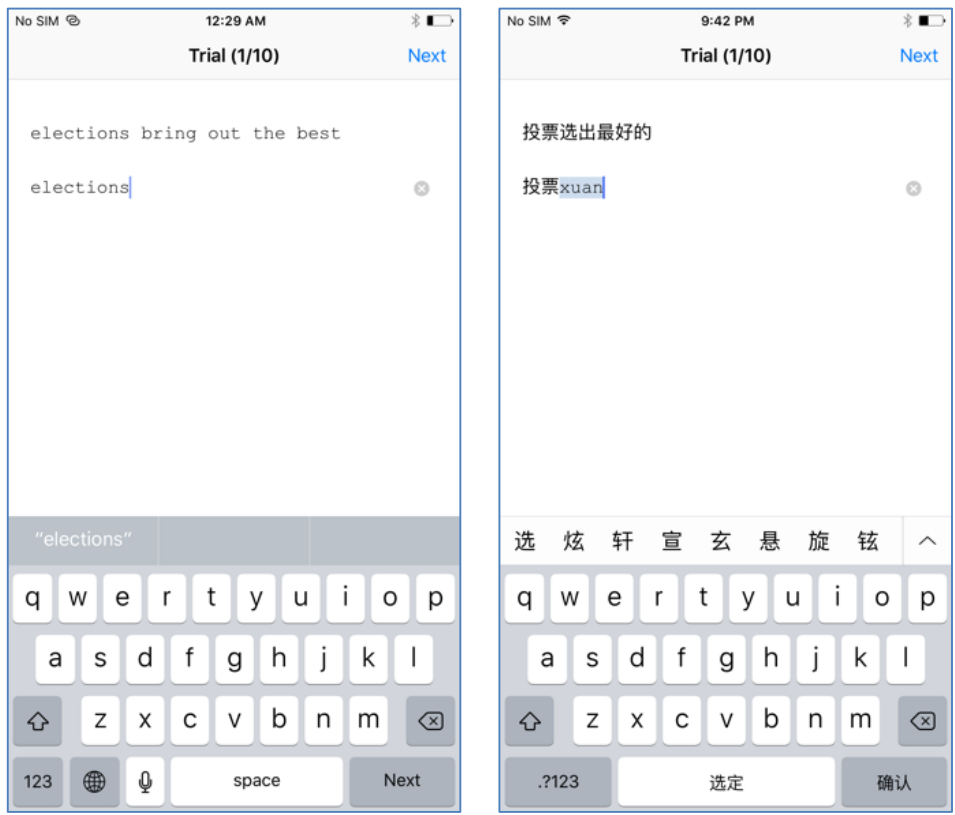

---

[2] We recognize, of course, that mobile text entry often takes place in noisy or distracting environments, possibly with the user in motion. Controlling for such factors is beyond the scope of this study, which, in seeking to establish an "upper-bound performance" for these two methods, chose to remove extraneous factors. After peak performance is rigorously established, subsequent studies can explore how extraneous factors degrade performance.

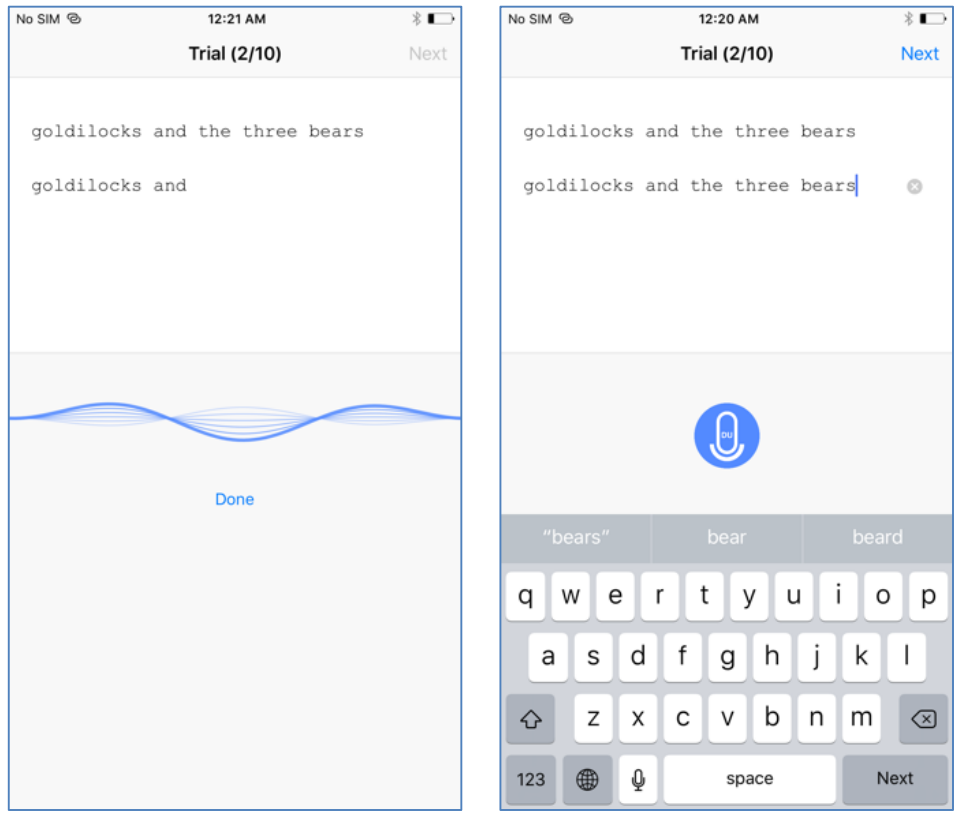

Fig. 1. **Test-bed application user interface. The phrase to transcribe is shown at the top of each interface with the user's input textbox just below. From left to right and top to bottom: (a) Keyboard input interface with English QWERTY keyboard and (b) Pinyin QWERTY keyboard. (c) Speech input interface: user is speaking and (d) after pressing "Done," a keyboard pops up for editing the initial spoken transcription, or speech can be used again.**

Fig. 1(c) and (d) show the two modes of our custom speech input method: speech recognition mode and an optional keyboard mode used for error correction. For the speech condition, the speech recognition interface is open at the start of each transcription task. The speech recognition system is on and the keyboard is hidden. The speech system recognizes the user's utterance and displays it in the textbox. To indicate that he or she is finished speaking, the user touches the "Done" button or touches anywhere on the screen, which switches the app to error correction mode. In error correction mode, the user can either touch the "mic" button to turn on the speech system again or correct errors using the keyboard. These designs were inspired by the user interface built into Apple iOS and are consistent with interfaces from prior studies of mobile speech recognition (e.g., [44]).

Although we chose to use the fastest typing-based keyboard we could find, we did not use advanced non-typing keyboard input methods such as stroke keyboards like Swype [64] or SHARK / ShapeWriter [20,55–57]. We omitted such input methods for a few reasons. For one, they are not on every phone. Market research shows that approximately 50M third-party keyboard apps had been cumulatively downloaded in the United States for Android [61] and iOS [60] as of mid-2016, while over 250M smartphones were in use in the U.S. at that same time (81% of 323M U.S. population owned smartphones [65]). These numbers indicate that at most 20% of American users employ these special keyboards. The actual usage is likely much smaller given multiple installations and the abandonment of downloaded apps after trying them. For another, such methods require initial practice to reach proficiency. In addition, prior studies have already established the performance of such methods, providing at least approximate figures with which to compare the results of this study. Those results showed that initial performance is about 15 words per minute, and after 40 minutes of practice, speeds reached about 25 words per minute with a 1.1% error rate [56].

All participants used the same iPhone 6 Plus that had our test-bed app installed. Since QWERTY is the most common keyboard layout, we used it as the default for both English and Mandarin speaking participants to increase ecological validity. We used Pinyin QWERTY as the default Mandarin Chinese keyboard as shown in Fig. 1(b). With a Mandarin Pinyin QWERTY keyboard, users input Mandarin characters by entering the Pinyin (phonetic transcriptions) of a Mandarin character, which triggers the presentation of a list of possible Mandarin characters matching the phonetic sound. Pinyin comprises the same 26 English letters displayed in the QWERTY layout. Using a QWERTY keyboard in this way is one of the most common ways people enter Mandarin Chinese text on smartphones and computers.

Because the speech recognition system already eliminates all invalid words before presenting a phrase, and because we wanted to test a state-of-the-art touchscreen keyboard with its associated features, it was reasonable to allow autocorrect in the English keyboard typing condition. The Mandarin Pinyin keyboard always outputs valid Mandarin characters, which can be regarded as having "implicit" autocorrect and spell check features. Chinese keyboards also have a built-in prediction feature that can provide the user with a collection of possible characters based on their previous input. As can be seen from Fig. 1(b), if the user presses the space bar, the first character "选" in the row right above the QWERTY keyboard will be selected. Likewise, in the English QWERTY keyboard setting, the space bar is used to confirm the auto-correct and spellcheck. Therefore, to remain consistent across languages and text entry methods, we enabled the standard Apple iOS text entry features, which are spell check, autocorrect, and word completion for both languages and both input methods.

### 3.3 Procedure

We had two equal phrase sets of 60 phrases each, called sets "A" and "B". (This division into sets was needed to avoid learning effects. For more on our phrases, see section 3.4, below.) With each text input method, participants transcribed 60 phrases drawn from one of the two phrase sets. One of the two phrase sets (A or B) was assigned in each session period in alternating order from experiment to experiment. Each participant entered text only in their native language, English or Mandarin Chinese, with each text entry method (keyboard or speech), the order of which was counterbalanced. The set of phrases used was the same for each language (translated, of course). Each phrase was regarded as one text entry "trial." For each text entry method, speech or keyboard, participants completed 10 practice trials before beginning the test, which itself consisted of 50 testing trials. Participants were taught how to use each interface before the study and during practice. This type of short (1 session) study is appropriate for the "walk up and use" standard of consumer systems: people are already familiar with speaking and they know how to type on a smartphone. Actions and timestamps were logged in the background (see section 3.5), and no timing information was ever visible to the participant. The test-bed app was set so that the appropriate text input method, language, and phrase set were selected prior to handing the phone to the participant. After entering all the phrases with both text input methods, the participant filled out a questionnaire regarding his or her demographic information and opinions of the two text entry methods. The phone was reset at the beginning of each experiment so that the phone would not learn better predictions based on previous users.

### 3.4 Text Transcription and the Phrase Set

For our experiment, we chose a text transcription (i.e., text copying) task rather than a text composition (i.e., text creation) task for a number of reasons. Prior studies [17,44] of mobile text entry, even speech-based text entry, have utilized transcription to separate the performance of the text entry methods, *per se*, from the writing abilities of the human participants. As others have argued [26], text composition introduces numerous confounds, such as thinking time, word choice, and challenges accounting for errors in view of unknowable human intentions. These confounds are serious obstacles to carrying out rigorous, measurable text composition experiments.

However, there are downsides to choosing text transcription that should be made explicit. For one, text transcription is less "naturalistic" than text composition. For another, it has been argued [39] that composing text with a manual keyboard interferes less with human thought than composing text with speech. And where thinking may give rise to speech disfluencies (e.g., "ah," "um"), speech recognizers can become confused—an irrelevant issue for manual input methods. As a result, using text transcription tasks to draw conclusions about speech recognition *vs.* touchscreen keyboards for general text entry might be suspect. That said, our endeavor here is to uncover the pure performance of the input methods themselves, providing an "upper-bound" for their speed and accuracy, understanding that text composition would reduce outcomes for both methods, perhaps differentially. Moreover, as the presented phrases for participants to transcribe become increasingly shorter, the differences between transcription and composition disappear, and speech disfluencies become much less of a concern. With our focus on short messages, some of these concerns are alleviated.

Based on the above rationale, we randomly selected 120 English phrases from a standard text entry phrase set of 500 phrases [27] to create our text entry phrase set. Our average English phrase contained 28.3 characters ($SD$=4.5), had a little capitalization (e.g., "Dow Jones Index," "Santa Claus," "Saturn"), and did not use punctuation. (James and Reischel [17] also used short messages that had no symbols, capital letters, or punctuation. And generating accurate punctuation using speech has been argued to be an unfair requirement for speech [5].) As ours was a general-purpose phrase set and not specific to mobile text entry, we compared it to findings from prior studies of mobile computing to ensure representativeness. For example, Feld et al. [10] reports on a micro-blogging SMS-based service with an average phrase length of 34.4 characters, or about seven words. Faulkner and Culwin [9] studied young adults' naturalistic text messaging behavior, finding a message was, on average, 11.05 words, or about 50 characters. Wood et al. [51] studied school children and college undergraduates text messaging behavior, finding that punctuation and capitalization were often omitted, the latter even at the beginning of sentences and for proper nouns. Thus, we find that our phrases were consistent with types of short messages occurring in everyday text messaging use.

We divided the 120 phrases in our phrase set into two equal sets, A and B, which were used for the two input method conditions to prevent learning effects. Both A and B used their first 10 phrases as a practice phrase set and the remaining 50 phrases as the test phrase set.

We also manually translated the 120 phrases into Mandarin Chinese[3] and used them as our phrases for the Mandarin part of the study. The phrases in the Mandarin set had a one-to-one correspondence to the phrases in the English set. They were typical of everyday Mandarin Chinese as well. The lengths of the English phrases varied from 16 to 37 characters ($M$=26.8, $SD$=4.3) and the lengths of Mandarin phrases ranged from 3 to 14 characters ($M$=7.7, $SD$=2.2).

## 3.5 Data Logging

Our test-bed app automatically logged all pertinent user behaviors, such as keystrokes, during the experiment. In addition, we logged timestamps with each of the actions listed below. During the study, a participant's actions fell into one of the following five categories. (The fifth item pertains only to the speech-based entry method.)

*Insert*. The user can insert a character using the keyboard or the speech recognition system. Characters added to the end of the current input stream are considered insertions as well, even though they are simply appended to the end.

*Delete*. The user can delete a single character using backspace, or multiple characters by selecting them first and backspacing, or delete everything simply by pressing the "X" button displayed at the end of the text box.

*Auto-Correct*. Auto-correct happens when a partial word or an existing word is replaced by a word suggested by the keyboard dictionary. The user presses the spacebar (i.e., continues typing) to confirm the auto-correct.

*Word Complete*. The user can insert multiple characters using the word completion feature. Word completion happens when the user selects a word from the suggested word list. This can happen when the user is at the beginning or in the middle of typing a word.

*Speech*. The speech system is turned on for the speech input method, but not for the keyboard input method. A speech "session" means a sequence of the following actions occurs in order: the user presses the "mic" button, the server starts to respond, the user starts to speak, the user stops speaking, the user starts to speak, the user stops speaking, …, the user presses the "Done" button, and the server finishes responding. We are able to timestamp each of these actions on the client application.

During a trial in the speech condition, the user starts entering text by speaking the presented string, after which they can correct it using either their voice or the keyboard. Hence, multiple speech sessions can be recorded for a single text entry trial (i.e., phrase).

---

[3] The English and Mandarin versions of the 120 phrases are available at http://hci.stanford.edu/research/speech/dataset.html so that other researchers can build upon our study.

## 3.6 Measures

We present and discuss the following empirical measures of text entry performance [48]. We use $T_i$ to denote the $i^{th}$ transcribed string, $P_i$ the $i^{th}$ presented string, and $S_i$ the $i^{th}$ phrase returned by the speech system prior to any edits made by the user (i.e., the initial speech transcription returned from the speech server).

*3.6.1 Words per Minute.* Words per minute is the most commonly used measure for text entry rates. The formal definition is given as follows: $WPM = \frac{|T_i|-1}{t_i} \times 60 \times \frac{1}{L}$, where $t_i$ is time in seconds for the $i^{th}$ trial. For the keyboard condition, it is measured from the entry of the first character to the entry of the last [25]. For the speech condition, time is measured from the onset of the first phoneme to the last edit made by the user. English words, by convention, are treated as having five characters [52], so we replace $L$ with 5.0. For Mandarin Chinese, $L$ was calculated as 1.5 because this is widely accepted as the average word length in Mandarin according to general statistics on the Chinese language [6].

Since we want to analyze not only the final transcribed string, but also what happens *during* the user's input, we need to define the "input stream" and classify characters therein [50].

The input stream can be logged as a sequence of strings. Table 1 shows the input stream of a trial with the presented string "where did I leave my glasses" in the speech condition. The first string is output by the speech system and the last string also represents the final transcribed string. By comparing two consecutive strings, we can see the user is making different types of changes (corrections): deleting the entire sentence using the X button, inserting a string using the speech system, deleting characters, or inserting characters. The actual actions that happened in the input stream are listed in the right column of the table. As we can see, these actions are either insertions or deletions. (Replacements can be treated as a deletion followed by an insertion). Note that users can insert or delete a single character or a whole string or substring.

*3.6.2 Error Rates.* We report text entry errors in two forms: uncorrected errors, which remain in the final transcribed string; and corrected errors, which are fixed (e.g., backspaced) during entry [42]. Correspondingly, we can have uncorrected and corrected error rates, which are normalized [0,1]. To compute these error rates, we need to classify each character in the input stream, including backspaces, into one of four character classes: *Correct*, *Incorrect-not-fixed*, *Incorrect-fixed*, and *Fixes* [42].

*Correct (C).* All correct characters in the transcribed text. The size of the class is computed as *MAX(P,T) – MSD(P,T)*, where *MSD* is the "minimum string distance" (also called the "edit distance") between two given strings [22,41,42,45]. Also, *P* and *T* stand for "presented" and "transcribed" string, respectively.

*Incorrect-not-fixed (INF).* All incorrect characters in the final transcribed text. The size of the class is computed simply as *MSD(P,T)* [42].

*Incorrect-fixed (IF).* All characters deleted during entry. The size of the class is computed as the sum of lengths of the value of all deletions.

Table 1. **An example input stream where the user transcribed "where did I leave my glasses" using speech input with keyboard-based error corrections.**

| Current Input Stream | Explanation | Action | Type | Value |
|---|---|---|---|---|
| wear did I live my glasses | Insert an initial string using speech | Insert | Speech | "wear did I live my glasses" |
| | Delete the entire string using X | Delete | Keyboard | "wear did I live my glasses" |
| wear did I leave my glasses | Insert a sentence using speech | Insert | Speech | "wear did I leave my glasses" |
| weardid I leave my glasses | Delete a space | Delete | Keyboard | " " |
| weadid I leave my glasses | Delete "r" | Delete | Keyboard | "r" |
| wedid I leave my glasses | Delete "a" | Delete | Keyboard | "a" |
| wdid I leave my glasses | Delete "e" | Delete | Keyboard | "e" |
| whdid I leave my glasses | Insert "h" | Insert | Keyboard | "h" |
| whedid I leave my glasses | Insert "e" | Insert | Keyboard | "e" |
| where did I leave my glasses | Select "where" from the predictions | Insert | Keyboard | "where" |

*Fixes (F).* All delete actions. Examples are deleting a single character with backspace or deleting the entire sentence with the "X" button at the end of the text box. Tapping the "X" is considered one "fix" action regardless of the length of the string in the text box.

With this classification, we are then able to compute the following measures for error rates [42,48] for both the keyboard and the speech condition: $uncorrected\ error\ rate = \frac{INF}{C+INF+F}$ and $corrected\ error\ rate = \frac{IF}{C+INF+F}$.

In addition, we can use the formula $utilized\ bandwidth = \frac{C}{C+INF+F+IF}$ to characterize the "efficiency" of an input method.

*3.6.3. Initial Speech Transcription Words per Minute.* Above, we computed the text entry rate in WPM using the final transcribed string. We can also calculate the speed of the initial speech transcription (IST) to estimate the initial text entry speed of the speech recognition system: $WPM_{IST} = \frac{|S_i|}{t_i'} \times 60 \times \frac{1}{L}$, where $t_i'$ is defined as the time between the user starting to speak and the server returning the last character; $S_i$ is the initial string returned by the speech recognition system before any error correction.

*3.6.4. Initial Speech Transcription Error Rate.* For the IST error rate, we can compare the IST string to the presented string. In this way, we can have a general sense of how accurate speech recognition is from the outset before any error corrections occur. We calculated the uncorrected and corrected error rates and the utilized bandwidth using the same formulae presented above; however, *IF* and *F* are always zero because we take the IST string generated by the speech system as the "final" string prior to any error correction.

*3.6.5. Other Speech-Specific Measures.* In addition, we present speech-specific measures that evaluate the general performance of the speech input method in terms of efficiency and effectiveness. These measures are intended to answer the questions shown in Table 2. These measures are novel and can be used by other researchers moving forward. Note that speech session time is defined as the entire time from when the phrase is shown to the participant until the speech system returns the final result. Server process time is defined as the time from when the user starts talking until the speech system returns the final result.

*3.6.6. Subjective Measures.* In addition to the performance measures, we asked participants to evaluate the workload of the tasks using the NASA TLX 7-point Likert scale across six categories [12,13]. Participants were also interviewed to gain an understanding of their subjective experiences using each input method.

Table 2. **Speech-specific measures and the questions that they answer. In the formulae, measures marked with an asterisk (*) can be computed in terms of time or in terms of number of characters.**

| Questions | Formula |
| --- | --- |
| How responsive is the server? | user talking time ÷ server process time |
| How responsive is speech input? | user talking time ÷ speech session time |
| How much delay is due to the user? | user delay time ÷ speech session time |
| How much delay is due to the system? | speech delay time ÷ speech session time |
| What percentage of time is spent on the speech system? | speech session time ÷ trial time |
| What percentage of time* is related to the keyboard? | keyboard time* ÷ trial time |
| What percentage of time is spent in inputting an initial speech phrase? | first speech session time ÷ trial time |
| What percentage of time* is used to make corrections? | correction time* ÷ trial time* |
| What percentage of correction time* is related to the keyboard? | correction using keyboard time* ÷ correction time |
| What percentage of correction time* is related to the speech system? | correction using speech time* ÷ correction time |

## 3.7 Design & Analysis

The experiment was a 2×2 mixed factorial design. The *input method* factor was a within-subjects factor and the *language* factor was a between-subjects factor. By designing our study as a factorial experiment, we were able to examine both main effects and interactions among languages and input methods. (We might, for example, discover that one input method is faster for one language but worse for the other.) Each experiment was divided into two "sessions," one session each for speech and the touchscreen keyboard. Each session consisted of practice and testing trials. One of the two phrase sets (A or B) was assigned in each session period in alternating order from experiment to experiment. The order of the sessions (speech or keyboard) was also counterbalanced across participants. Our analysis of variance shows that neither *gender* nor *input method order* exhibited a main effect for any measure, the latter indicating that our counterbalancing worked.

Every keystroke was recorded together with a timestamp. There were 48 log files, one per participant, and each contained 50 testing trials for the speech input method and 50 testing trials for the keyboard input method. Hence, we had 4800 data points from the study. In our statistical analyses, we kept all individual trial-level measures. For the Pinyin QWERTY keyboard, we recorded both letters (Pinyin) and the resulting Mandarin Chinese characters. The information was logged as a JSON file during the study. We wrote a parser for the auto-generated log file and an analyzer to calculate the aforementioned series of text entry measures such as entry rates and error rates.

We ran appropriate statistical tests in **R** on the processed data for analysis. Specifically, we analyzed words per minute (WPM) using a parametric mixed-effects model analysis of variance [11,23,46], with fixed effects of *input method* and *language* and a random effect of *participant*. We analyzed error rates and other data using the nonparametric Aligned Rank Transform procedure [14,36,49]. Quartile-quartile plots of the fitted residuals of our statistical models show that residuals were normally distributed, indicating the appropriateness of our models for analyzing our data.

## 4 RESULTS

In this section, we present the results of our study of smartphone text entry using both a touch-screen keyboard and speech recognition. We do so for two languages, English and Mandarin Chinese.

### 4.1 Participants' Familiarity with Smartphones

All participants were familiar with typing on an Apple iPhone. As it will be seen, average typing speed in English for our study was about 52.2 words per minute; prior work shows averages around 41 WPM for transcription [66], indicating our participants might have been experienced smartphone users. Similarly, average typing speed in Mandarin Chinese for our study was about 42.8 WPM; prior work shows averages around 27 WPM for transcription [67]. It is safe to conclude that our participants were generally proficient at entering text on mobile phones.

In addition, we surveyed participants on their usage of their own smartphones and the results are summarized below in Fig. 2 and Fig. 3.

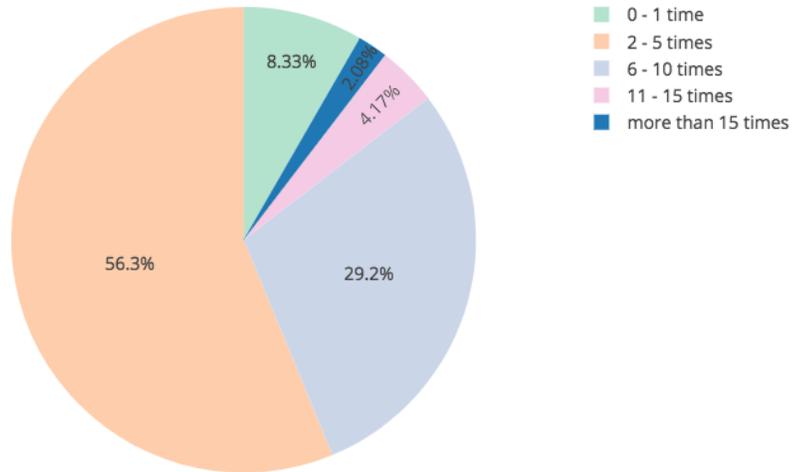

Fig. 2. **Distribution of number of times per hour participants reported interacting with their smartphones.**

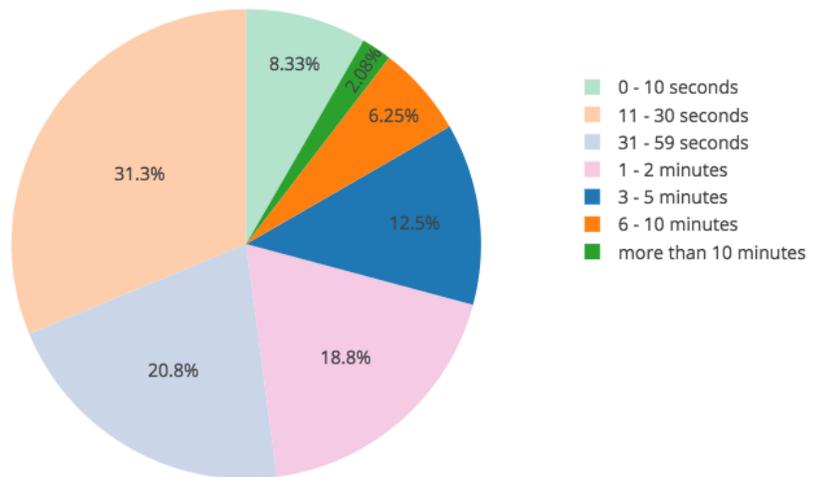

Fig. 3. **Distribution of participants' reported duration of each use of their smartphones.**

For each user, we multiplied their number of times per hour spent on a smartphone with the estimated duration of each use to obtain their average self-reported time spent on smartphones every day. Then for each input method, we calculated a simple linear regression to predict WPM based on daily phone use time. Our results show that there was no significant relationship found for either the speech input method rate ($F_{1, 46}$=0.21, $p$=.65 with an $R^2$ of .005) or the keyboard input method rate ($F_{1, 46}$=0.048, $p$=.83 with an $R^2$ of .001).

We also collected data on which tasks participants do weekly using their smartphones. Their results are shown in Fig. 4.

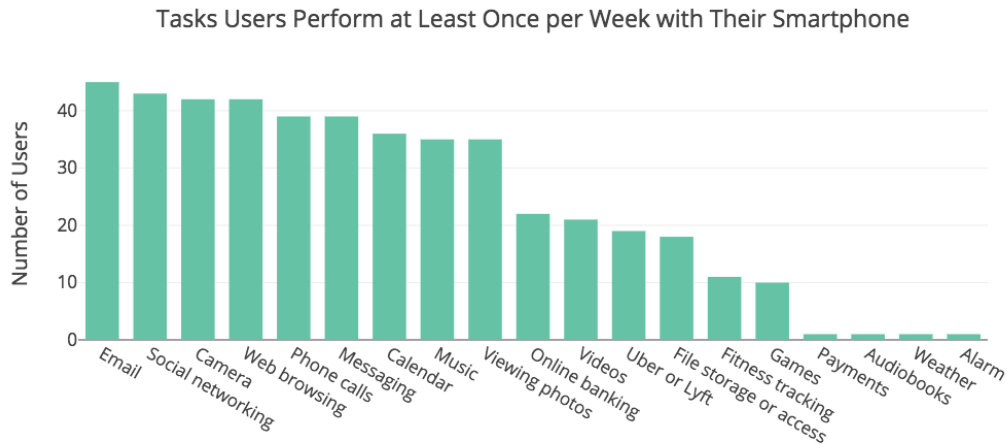

Fig. 4. **Tasks participants say they perform at least once per week with their smartphones.**

## 4.2 Speed

Means and corresponding standard deviations (in parentheses) of six measures are shown in Table 3. Text entry speeds in words per minute (WPM) for both methods in both languages are graphed below in Fig. 5 and listed below in Table 3. A mixed-effects model analysis of variance indicates that both *language* and *input method* exerted a significant main effect on entry speed: $F_{1,65}$=16.00, $p$<.001 for *language*, and $F_{1,196}$=1182.39, $p$<.001 for *input method*. Moreover, there was a significant interaction between these two factors: $F_{1,196}$=15.12, $p$<.001. The significant interaction can be seen in the interaction plot in Fig. 5. One can see that English speed increases more than Mandarin speed when moving from the keyboard to speech. We conducted *post hoc* pairwise comparisons corrected with Holm's sequential Bonferroni procedure [16]. These tests show that all pairwise comparisons between levels of *language* and *input method* are statistically significant. In conclusion, speech is faster than keyboard. Also, although our presented phrases expressed the same meanings in both languages, English entry rates were significantly faster than Mandarin for both keyboard and speech. (For more discussion on comparisons between Mandarin and English, see section 5.)

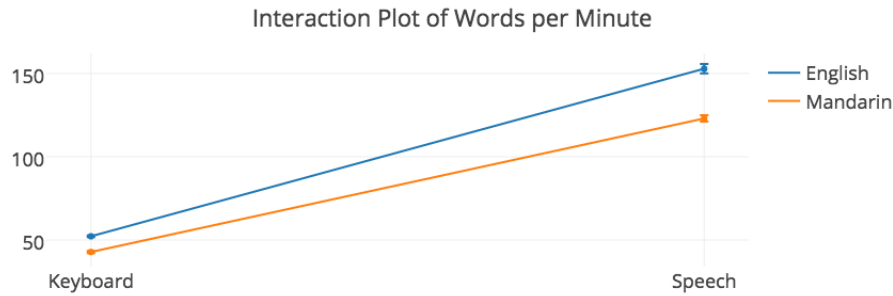

Fig. 5. **Words per minute by language and input method. Higher is faster (better). Error bars represent +/-1 standard error.**

The average entry rate of speech input was 2.93 times faster than keyboard input for English, and 2.87 times faster than Mandarin Pinyin input. Moreover, the initial speech transcription (IST) was 3.42 times faster than keyboard for English and 4.32 times faster for Mandarin, offering an estimate of the speed of "perfect speech" requiring no error correction. The ratio of average IST speed to average speech input speed was 1.17 for English and 1.50 for Mandarin, which characterizes how much entry rate speed (in WPM) error correction of the IST is costing users. If the IST were always perfect, then $WPM_{IST}$ would equal the speech entry rate, and this ratio would be 1.0.

Table 3. **Mean and corresponding standard deviation (in parentheses) of five measures for the two input methods and the initial speech transcription (IST) before error correction, for both languages.**

|  | English | | | Mandarin Chinese | | |
|---|---|---|---|---|---|---|
| Input Method | Keyboard | Speech | IST | Keyboard | Speech | IST |
| Words per Minute | 52.24 (17.19) | 152.86 (98.14) | 178.92 (45.89) | 42.83 (19.70) | 123.00 (66.36) | 184.97 (119.65) |
| Uncorrected Error Rate | 0.35% (2.00%) | 0.55% (3.63%) | 4.37% (12.28%) | 1.22% (5.41%) | 2.06% (6.27%) | 9.94% (20.48%) |
| Corrected Error Rate | 4.72% (8.31%) | 3.93% (12.25%) | n/a | 17.73% (22.98%) | 6.67% (15.81%) | n/a |
| Utilized Bandwidth | 91.69% (13.11%) | 94.46% (14.39%) | 95.63% (12.28%) | 75.14% (28.05%) | 89.10% (20.20%) | 90.06% (20.48%) |
| Auto Correct Portion | 50.00% (76.73%) | 1.17% (12.37%) | n/a | n/a | n/a | n/a |

We also observed a larger standard deviation of entry rate for speech input compared to keyboard input for both languages (98.14 vs. 17.19 WPM for English, and 66.36 vs. 19.70 WPM for Mandarin). As with human performance tasks in general, further practice improves speeds and reduces performance variance [4], suggesting that more practice might improve speech-based input speeds and reduce this standard deviation further.

### 4.3 Error Rates

Uncorrected and corrected error rates are shown above in Table 3. We used the nonparametric Aligned Rank Transform procedure (ART) [14,36,49] to analyze two error rates—uncorrected error rates and corrected error rates—because error rates do not meet the assumptions for parametric analyses. Uncorrected error rates characterize the portion of errors that are not rectified by users, such as omissions, insertions, and substitutions. Results show a main effect of *language* ($F_{1,162}$=11.80, $p$<.001) and of *input method* ($F_{1,196}$ = 9.00, $p$<.01) on the uncorrected error rate, and of their interaction ($F_{1,196}$=9.12, $p$<.01) as illustrated in Fig. 6. Speech is slightly more prone to leaving uncorrected errors than the keyboard, especially in Mandarin. This finding represents a slight speed-accuracy tradeoff because speech is also faster than the keyboard, as described above.

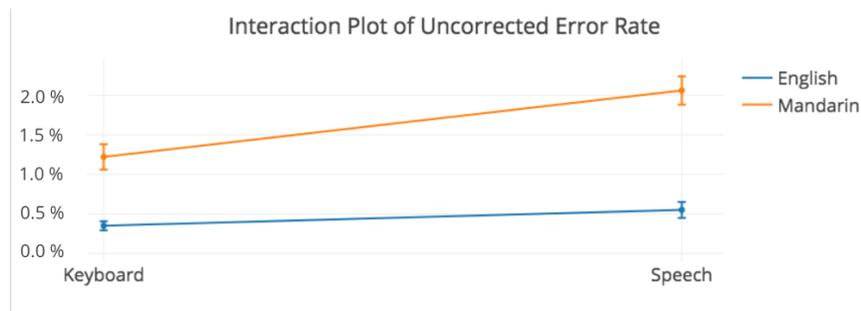

Fig. 6. **Uncorrected error rate as a function of language and input method. Lower is more accurate (better). Error bars represent +/-1 standard error.**

For corrected error rates, which refer to errors made but corrected during input, we found a detectable difference for *input method* ($F_{1,196}$=198.26, *p*<.001), *language* ($F_{1,55}$=44.55, *p*<.001), and their interaction ($F_{1,196}$=110.43, *p*<.001). These findings indicate that users spent significantly more time and effort editing their text when using the keyboard than when using speech. The difference is more pronounced in Mandarin than in English (6.67% vs. 17.73% in Mandarin, and 3.93% vs. 4.72% in English). One possible reason for this is that users have to press multiple keystrokes to enter a Pinyin (phonetic transcription) for a Mandarin character, while only a single keystroke is needed for an English character. From the interaction plot shown in Fig. 7, it is clear that the keyboard makes more errors during entry than speech, and this difference is much greater in Mandarin than it is in English.

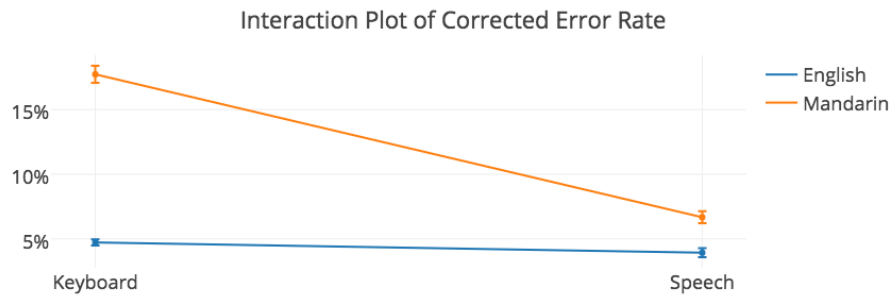

Fig. 7. **Corrected error rate as a function of language and input method. Lower is more accurate (better). Error bars represent +/-1 standard error.**

Unlike for WPM and uncorrected error rate, where *post hoc* contrast tests were non-significant, corrected error rate contrasts were statistically significant. Specifically, we use interaction contrasts [2,28] for cross-factor interaction tests, and the results reveal a significant difference ($\chi^2_{(1,N=4800)}$= 1075.7, *p*<.0001). In Mandarin, there is a significant and greater difference in the keyboard versus speech corrected error rate than in English, which we can see from Fig. 7. The significant interaction contrast justifies the use of further nonparametric contrasts. For each participant, we averaged their speech and keyboard corrected error rates across their trials, and compared them within each language using a Wilcoxon signed-rank test [47] corrected for multiple comparisons [16]. The test showed that in both English and Chinese, there was a statistically significant difference in corrected error rates between speech and keyboard (*p*<.001).

To summarize, for errors made and fixed during entry, the keyboard was much less accurate than speech (11.22% vs. 5.30% corrected error rate), especially in Mandarin. However, speech left more errors in the transcribed text than the keyboard did (1.30% vs. 0.79%). Therefore, speech made fewer errors during entry but left a few more errors after entry. This finding might be because users were less willing to correct errors with the keyboard in a hands-free setting, or because keyboard users see each word as it is typed and so are more likely to notice errors. Follow-up studies with a range of speech-based user interfaces to further explore this result would be interesting.

### 4.4 Utilized Bandwidth

Utilized bandwidth is a measure that characterizes the proportion of keystrokes that contributed to correct parts of the final transcribed string. We utilized the nonparametric Aligned Rank Transform procedure to examine the effects of *input method* and *language* on utilized bandwidth. Our results reveal a statistically significant effect of *language* ($F_{1,60}$=51.08, $p<.001$), *input method* ($F_{1,196}$=134.61, $p<.001$), and their interaction ($F_{1,196}$=63.15, $p<.001$). An interaction plot is shown in Fig. 8. An interaction contrast [2,28] shows that the difference between keyboard and speech is significantly greater in Mandarin than in English ($\chi^2_{(1,N=4800)}$=942.35, $p<.0001$). Such an interaction contrast warrants using additional nonparametric contrasts to examine pairwise comparisons. A Wilcoxon signed-rank test [47] corrected for multiple comparisons [16] comparing speech to the keyboard in English was statistically significant ($p<.001$), and the same for Mandarin ($p<.001$). Therefore, speech uses input bandwidth more efficiently than the keyboard in both Mandarin and English.

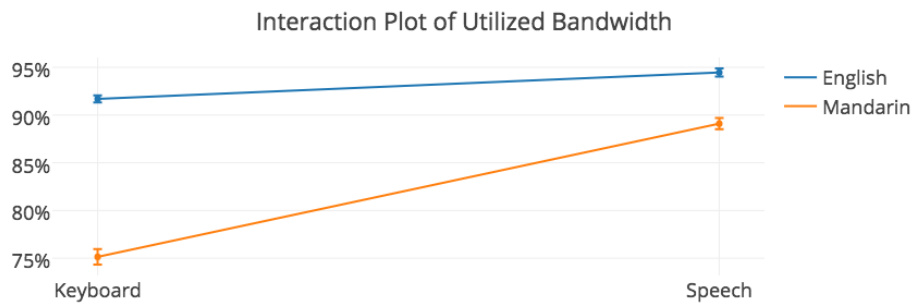

Fig. 8. **Utilized bandwidth as a function of language and input method. Higher is more efficient (better). Error bars represent +/-1 standard error.**

### 4.5 Auto Corrections and Word Completions

Finally, we present the total occurrences of auto-corrections and word completions over 50 trials. As seen in Table 3, occurrence of these features was higher for the keyboard than for speech in English. These features were not explicitly supported for the Pinyin QWERTY keyboard, as the nature of Pinyin is to show only valid words anyway. Therefore, *input method* becomes the only independent variable and auto-corrections is the dependent variable. Using the Aligned Rank Transform procedure for nonparametric analysis of variance [14,36,49], we find there was a significant effect of *input method* ($F_{1,196}$=532.01, $p<.001$) on the number of auto-corrections. Specifically, many more auto-corrections occurred when participants used the keyboard input method (50.00%) than when they used the speech input method (1.17%).

### 4.6 Speech-Specific Results

We present the following speech-specific results to assess the performance of the speech-based input method in many respects. Fig. 9 shows the time proportion participants spent on different tasks when entering text. As can be seen, only slightly more than one third of the total time (39.2%) was needed for users to actually speak a phrase. This result can be explained in part by the short lengths of the presented phrases, which were similar in length to short text messages. A sizeable amount of time (31.5%) was wasted due to users' delay. For example, many users did not start to speak until one or two seconds after they turned on the speech system. Also, several users hesitated to press the done button even after finishing speaking, and they explained that this was because they were not sure if they wanted to talk more. This behavior could lead to a substantial delay since the speech recognition system is context based [1], i.e., it does not finalize the output until the user clearly indicates that he or she is finished speaking. All these waiting activities constituted a considerable user delay. This finding also suggests that as users gain experience with speech input, they can significantly reduce user delay, thereby allowing speech to gain a further speed advantage. In contrast, less delay was due to the speech recognition system (22.6%). This delay mainly occurred when the system was still processing the user's utterance after the "Done" button was pressed. These statistics can also be useful to builders of speech systems, as they can see where to improve overall performance.

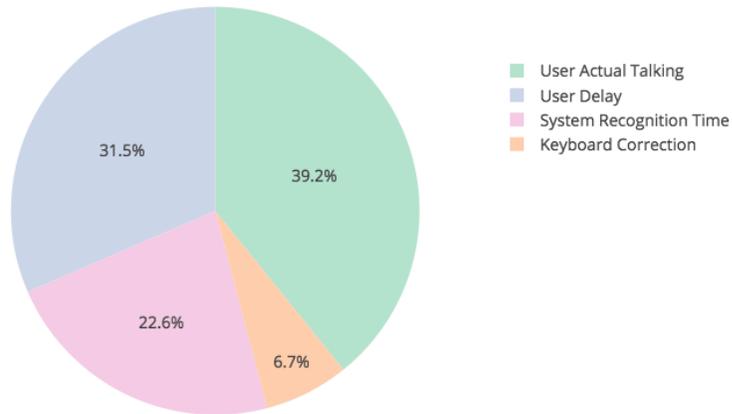

Fig. 9. **Time proportion of different users' actions and recognition processing time with the speech input method.**

Fig. 10 shows the time the user spent on speech input and keyboard input (error correction mode) in completing tasks using the speech input method. Users spent substantial time (91.5%) in producing the initial transcription using the speech recognition system. Participants used the rest of the time (8.5%) to correct errors; of that, 86.0% of the time users corrected errors using the keyboard and only 14.0% of the time speech input was used.

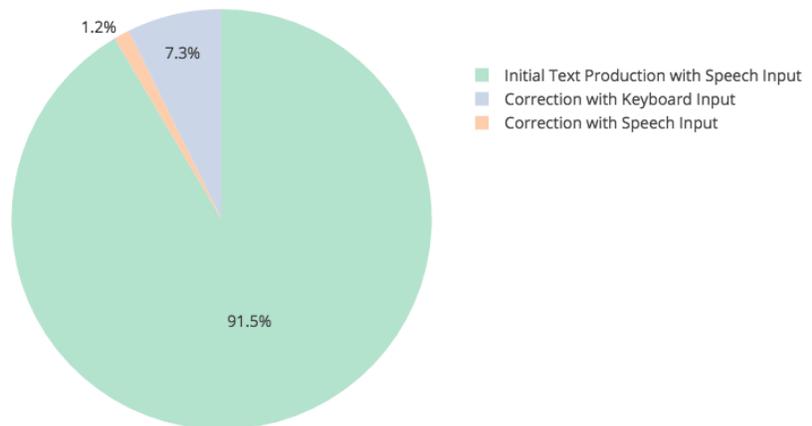

Fig. 10. **Time proportion of different stages with the speech input method.**

We can also view the same splits in the number of characters rather than time. As can be seen from Fig. 11, 96.7% of the final characters were generated from the initial speech input directly. A much smaller 2.7% and 0.6% of the final characters resulted from corrections using the keyboard and speech, respectively. These proportions suggest that users' confidence in the speech system's initial result was high enough that they barely resorted to correction mode to modify the initial input. Furthermore, participants preferred to correct errors using the keyboard instead of speech most of the time. Our post-experiment interviews indicate that participants found it more efficient and comfortable to correct errors with the keyboard than with speech.

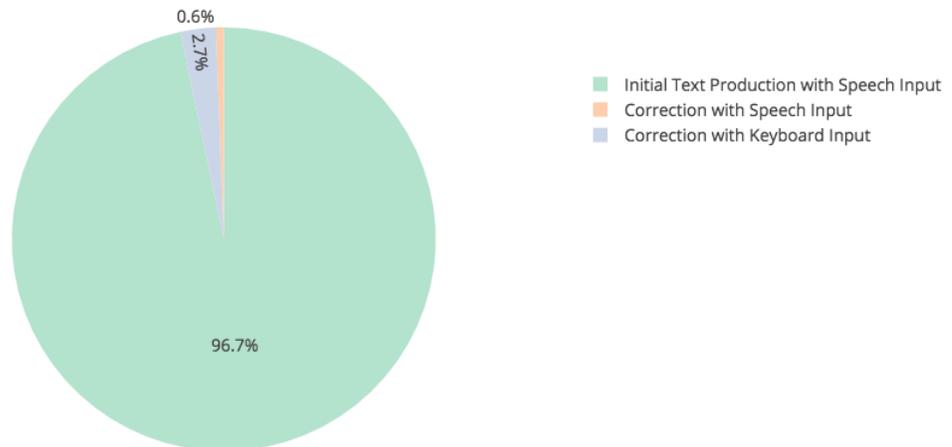

Fig. 11. **Character proportion in tasks with the speech input method.**

### 4.7 Subjective and NASA TLX Ratings

We collected participants' subjective ratings and NASA TLX ratings on their perceived demand, effort, frustration, and performance of the two input methods through paper-based questionnaires. Fig. 12 and Fig. 13 summarize the results. The lower the numbers are, the less demanding or the better performance the user found the method. As illustrated by Fig. 12, participants found, for both languages, the keyboard easier to correct errors with, but speech input easier to produce text.

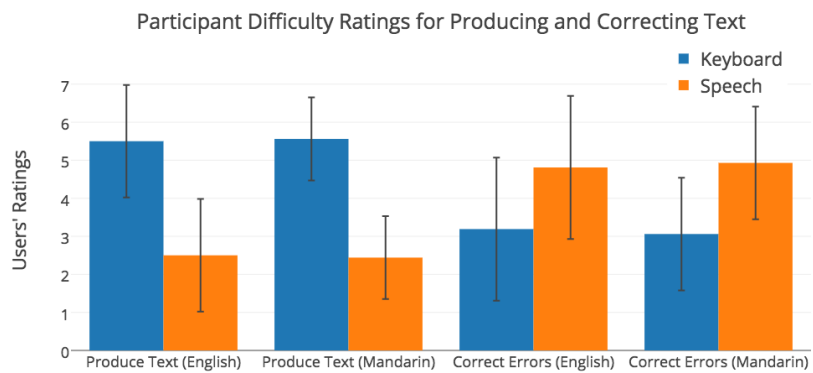

Fig. 12. **Participants' subjective evaluation of difficulty. Ratings were on a 1-7 scale. Lower is less difficult (better). Error bars represent +/-1 standard deviation.**

159:18 • S. Ruan et al.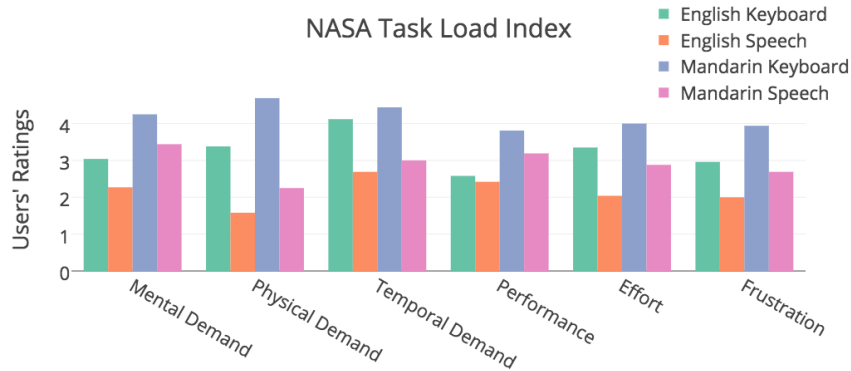

Fig. 13. **Participants' ratings on the NASA TLX workload instrument. Ratings were on a 1-7 scale. Lower values correspond to less demand, better performance, less effort, and less frustration for their respective scales. In general, speech input was perceived as having a lower workload than keyboard input.**

On the NASA TLX 1-7 scale shown in Fig. 13, English speech was ranked as the easiest input method among all four options across all six categories, and the Mandarin Chinese keyboard was ranked the most difficult. Speech was rated as having lower workload than the keyboard in all six categories for both English and Mandarin.

We averaged all ratings across six categories and again utilized the Aligned Rank Transform procedure [14,36,49] to conduct a nonparametric factorial analysis of variance. Results show that *language* ($F_{1,40}$=12.67, $p$<.001) and *input method* ($F_{1,40}$ = 32.63, $p$<.001) exerted a statistically significant effect on ratings, but not their interaction ($F_{1,40}$=0.28, *n.s.*). Hence, speech is, in general, less demanding than the keyboard, and English transcription tasks are less demanding than Mandarin transcription tasks. An interaction plot is shown in Fig. 14.

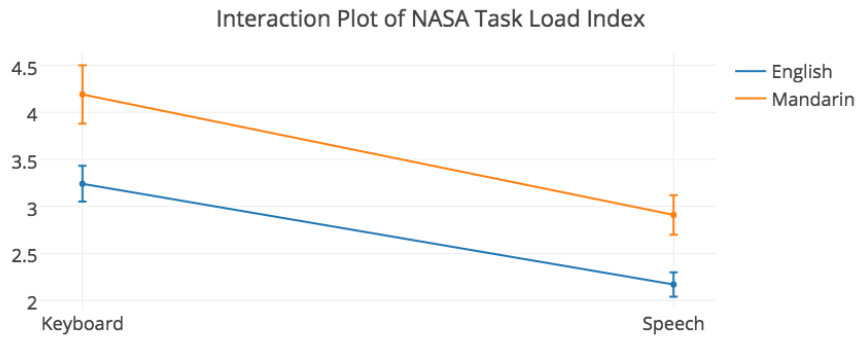

Fig. 14. **NASA TLX ratings by language and input method. Ratings were on a 1-7 scale and averaged across all six NASA TLX scales. Lower is lesser workload (better). Error bars represent +/-1 standard error.**

## 4.8 Interview Results

We also interviewed all participants for their qualitative opinions of the two input methods. They expressed that speech methods were generally easier to grasp and use, less demanding, and more accurate than their original expectation of speech recognition.

Most users expressed that they liked the coherence and continuity of the speech system compared to keyboard input: "It is difficult to type an entire sentence using the keyboard, since a typo at the beginning of the sentence could be very hard to correct later." Users' attitudes towards keyboard input were that it was more comfortable to use but also more error-prone: "I am comfortable typing on a keyboard. There was less uncertainty about what text was going to be produced. I could correct errors as they were made. That being said, it seemed like I introduced more errors typing than the speech system did."

Despite these advantages of speech input, users also indicated some deficiencies with speech recognition: "When I made mistakes, it seemed like it took longer to correct, because I was switching from holding the phone to speak vs. holding to type."

Our Chinese participants also provided some interesting insights regarding Mandarin Chinese input. Some stated that the Mandarin Pinyin keyboard was inherently ambiguous because a single Pinyin can map to many different characters that have the same phonetic sound and selecting the right one took some effort. Participants pointed out speech could help address the problem of having to come up with a close-to-correct spelling for a word: "Compared to the speech method, [when using the keyboard] I also had to worry about the 'rough' spelling of the word, which took more effort." Some Mandarin speaking participants commented that they liked the flexibility and the ability to recognize accents with the Mandarin speech recognition system, especially when they were not sure about the exact Pinyin (phonetic sounds without any accent), which is usually required to produce the right characters. There were also challenges specific to Mandarin recognition: "Some phrases in Mandarin sound the same, so if the speech recognition software does not have the context, it is very hard for it to guess the correct output."

## 5 DISCUSSION

In this section, we review our major findings, and reflect on our study limitations and broader implications of this work.

### 5.1 Quantitative Analysis

Our results revealed that the speech input method was faster (2.93 times faster in English, and 2.87 times faster in Mandarin Chinese) than keyboard input. English has higher input rates than Mandarin in terms of both keyboard (52.24 vs. 42.83 WPM) and speech (152.86 vs. 123.00 WPM). Considering that our English and Mandarin phrases expressed the same content, bilingual speakers could possibly experience a similar speedup in both languages when using the speech input method.

Speech input was less error-prone than keyboard input *during* entry, but slightly more prone to leaving uncorrected errors *after* entry. This tradeoff might suggest that people are more attentive to their text when using keyboard input because of the way in which speech recognition worked in the study. It was practically infeasible to correct speech recognition errors until the entire phrase had been dictated. With keyboard entry, users could presumably choose between correcting errors as soon as they noticed them and correcting all errors in a separate pass after entering all of the text.

We found that the Deep Speech 2 engine's initial speech transcription could itself achieve a respectable 4.37% uncorrected error rate in English, and a 9.94% uncorrected error rate in Mandarin. Therefore, users only needed to make a few corrections when using speech input. The initial recognition rate for English and Mandarin Chinese were both rather remarkable, especially considering that even humans do not demonstrate perfect speech recognition [24].

### 5.2 Qualitative Analysis

Participants' subjective ratings and testimonials revealed their preference for the speech input method over the keyboard. In general, they found speech to be more natural, smooth, and capable of recognizing most of their words immediately. Moreover, the speech-based input method did not require much practice for novice users. The practice set consisted of 10 phrases and participants could opt to practice with that set many times until becoming ready for the test. In our study, however, participants became quite accustomed to speech after only 10 practice phrases, and no participants chose to undergo any additional practice.

### 5.3 Potential for Improvement

The high speed of speech input was achieved due to the responsiveness of the speech recognition system, as demonstrated by our results showing that speech processing time constituted only 22.6% of the entire experiment time. By contrast, the user delay (e.g., after finishing talking but before tapping the "done" button) was 31.5% of the experiment time. Our speech-specific measures and the corresponding results give designers insights into further improvements. The speech input speed might improve if users reduce this delay through experience with the system, or if we can create a design that helps even novice users reduce this delay. For example, we might be able to create a new interface that encourages users to signal the speech system as soon as they are finished speaking. Rather than having the user manually specify when they are done talking, we also think auto-detecting the end of speech holds significant promise. It also relieves the user of another manual step.

Another improvement would be the incorporation of customizable system features into the speech recognition system itself. Over the course of 50 trials, on average, auto-corrections and word completions occurred in exactly half of the trials (25.0 times) in the English keyboard condition, but barely appeared (0.6 times) in the English speech condition, as shown in Table 3. We could envision that a speech system with support of system features would be more powerful and desirable. For example, one design could be a system that upon selecting a word for correction gives the most likely words based on computed context-based probabilities (i.e., an "*n*-best list") so that the user could select the right word from the list instead of typing it out. Moreover, the speech recognition system could analyze and learn from the word corrections made by the user. With the data collected, the system could keep updating the underlying machine learning model to make more accurate and customized predictions for each user.

### 5.4 Limitations and Future Work

As stated at the outset of this paper, we sought to obtain an "upper-bound performance" measure for both speech and keyboard text entry under ideal conditions. Speech was no doubt aided by a quiet environment and fast network speeds. The keyboard was aided by having participants seated and not walking, and by the high familiarity participants already had with mobile keyboard-based text entry, which they generally did not have with speech-based systems. Having now provided such a performance measure for two languages, we think that modern speech systems like Baidu's Deep Speech 2 are promising enough to warrant further investigation, especially under less-than-ideal real-world settings. For example, a study of speech input under a variety of noise conditions would be interesting.

We only examined a server-based speech recognition system in this study. Embedded speech recognition systems are less common but have lower latency and better reliability [29]. As more research on how to improve accuracy and lower memory consumption develops, an evaluation of embedded speech applications is warranted.

Following other studies of mobile text entry (e.g., [17]), we also only tested transcription tasks on short messages without punctuation. But in reality, people need to use speech functionality in a variety of ways, including to enter punctuation and symbols. Future work should investigate both explicit and implicit methods for generating punctuation and symbols with speech. With a more fully developed speech recognition technique, one including punctuation and symbols, we could design an effective email composition tool, for example.

There have been other keyboards that have been developed as alternatives to the standard built-in Apple iOS (and Google Android) keyboards, including swiping keyboards, and specialized Pinyin keyboards such as the Baidu IME. These keyboards require some substantial user training, unlike the speech interface where all participants familiarized themselves with the system after only 10 practice phrases. Other research showed that there was no significant difference in entry rates between Swype and QWERTY, but Swype was rated higher in subjective rankings [30]. Another study showed that with 13-19 hours of practice, users' typing performance using KALQ could be 34% more efficient than typing on split-screen QWERTY layouts [31]. A detailed comparison of speech-based input to these keyboards, especially for highly trained users, would also be interesting.

## 6 CONCLUSION

In this research, we designed and analyzed a state-of-the-art speech input method for mobile devices in English and Mandarin Chinese, and compared its performance to a state-of-the-art touchscreen keyboard in a controlled laboratory text entry study. In general, we found that speech was nearly three times faster than the keyboard and made fewer errors during entry, but left slightly more errors after entry was complete. We provide suggestions for how to further optimize speech input using the results obtained from the study. This work contributes the first empirical study demonstrating the performance of speech input compared to keyboard input on touchscreen phones under ideal conditions for both state-of-the-art text entry methods. Our work also provides novel speech-specific text entry measures to evaluate the performance of speech input. These results offer a baseline to which future "in the wild" studies can be compared. We hope our results will attract more researchers to develop effective speech-based mobile applications and interfaces in the near future.

## ACKNOWLEDGMENTS
The authors thank He Dang of Baidu's Speech Technology group, who performed a preliminary study that inspired this research project.